\pdfoutput=1
\documentclass[reprint, aps, prd, preprintnumbers, amsmath, amssymb, superscriptaddress, nofootinbib]{revtex4-1}
\usepackage{amssymb,amsmath,bbold,slashed}
\usepackage{graphicx,soul}
\usepackage[usenames,dvipsnames]{xcolor}
\usepackage[unicode=true,pdfusetitle,bookmarks=true,bookmarksnumbered=false,bookmarksopen=false,citecolor=Turquoise,
 breaklinks=false,pdfborder={0 0 1},backref=false,colorlinks=true,pdfpagemode=FullScreen]
 {hyperref}

\begin{document}

\title{Electric Dipole Moments From Missed Dark Matter Scattering}

\author{Jason L. Evans}
\email[]{jlevans@sjtu.edu.cn}
\affiliation{T. D. Lee Institute and School of Physics and Astronomy, Shanghai Jiao Tong University, Shanghai 200240, China}

\begin{abstract}
 Axion-like particles are a well-motivated candidate for ultralight dark matter. Because dark matter must be non-relativistic, the effects of its scattering with Standard Model particles are negligible and generally go unnoticed. However, due to the large occupation number of ultralight dark matter, the sum of all scatterings leads to a classical field-like interaction with Standard Model particles. In the case of an axion-like particle, this scattering imparts a parity violating effect. If this collective scattering with axion-like particles is inserted into the one-loop quantum electrodynamics diagram, the parity violation imparted by this scattering will convert the anomalous magnetic moment contribution into an electric dipole moment. This contribution is quite large and leads to a prediction inconsistent with precision measurements of the proton and electron electric dipole moments, unless their couplings to the axion-like particles are very weak. As a result, the constraints on the couplings of axion-like particle dark matter to the electron and proton are improved by as much as eleven and six orders of magnitude, respectively. 
\end{abstract}

\maketitle

\section{Introduction}
One of the great puzzles of particle physics is the nature of dark matter. As of yet, we know little about its mass and couplings, and still do not even know if it is bosonic or fermionic in nature. This lack of understanding about the nature of dark matter is further exacerbated by the fact that theoretically motivated candidates, like the weakly interacting massive particle\cite{Goldberg:1983nd,Ellis:1983ew}, have been challenged by experimental searches.  

Other well motivated dark matter candidates\cite{Preskill:1982cy,Abbott:1982af,Dine:1982ah,Arias:2012az}, like the QCD axion\cite{Peccei:1977hh,Kim:1979if,Shifman:1979if,Dine:1981rt,Zhitnitsky:1980tq} , were conceived of several decades ago but have evaded all our best experimental searches. Although progress is being made and planned experiments give us hope of reaching the predicted parameter ranges, detection of QCD axion dark matter is still an elusive quest.  

Axion-like particles\cite{Svrcek:2006yi,Arvanitaki:2009fg}, a generalization of the QCD axion, are another dark matter candidate motivated by string theory. While these types of particles have less theoretically constrained couplings, much of their parameter space is challenging to probe and remains relatively unexplored by current experiments. In fact, masses in the range ${\mathcal O}(10^{-20}~eV )-{\mathcal O}({eV})$, with couplings as large as ${\mathcal O}(10^{-10})$, are still allowed. 

In spite of the challenges, searches are on-going, with many different experimental programs aimed at searching for axion-like particles that couple to the Standard Model. The current best constraints on these couplings are mass dependent and come from a myriad of experiments. The photon coupling to the axion is constrained by many different experiments with the most significant ones coming from ADMX\cite{ADMX:2024xbv}, which can constrain some of the QCD axion window. The strongest constraints on the axion-electron coupling come from red giant observations\cite{Capozzi:2020cbu}. If the axion-like particle (ALP) is dark matter, the electron ALP coupling is further constrained in the low mass range by $g_e-2$ experiments, due to the background enhanced propagation of the ALP \cite{Arza:2023wou,Evans:2024dty,Evans:2023uxh}. A small portion of the larger end of the mass range is also constrained by either the XENON experiments \cite{XENON:2019gfn,XENON:2020rca,XENON:2022ltv} or searches for anomalies in X-rays\cite{Ferreira:2022egk}. The axion-proton coupling constraints come from SN1987 observations\cite{Lella:2023bfb}.  




Searches for ultralight dark matter tend to be challenging, due to the tiny recoil energy of interactions with the dark matter. For much of the viable parameter space, the recoil energy is well below the energy resolution of current experiments. However, the large occupation number of an ultralight dark matter background tends to act like a (quasi-)coherent state and collectively can lead to an observable effect. Previous work has examined the effect of this (quasi-)coherent state on the electron electric dipole moment (EDM). In \cite{Hill:2015kva,Hill:2015vma,Hill:2015lpa}, it was shown that the axion-photon coupling induces an EDM for the electron through a photon propagator \footnote{However, this paper's results have faced significant skepticism \cite{Flambaum:2015ica}.}.  Work on examining the effect of the ALP electron coupling on the electron EDM was first done here \cite{Alexander:2017zoh}, where, the non-relativistic limit of the Dirac equation was taken for a non-derivative coupling of the electron to the ALP. It was claimed that this leads to a rather large EDM, which was further discussed in \cite{Chu:2019iry,Wang:2021dfj,Smith:2023htu}.  However, several ensuing works \cite{DiLuzio:2023ifp,Berlin:2023ubt} have since argued that this contribution is likely fictitious. The key indicators of a fictitious contribution were a lack of concordance between the results for the derivative and non-derivative coupling, no higher order contributions in the non-relativistic limit consistent with a relativistic origin, and lack of shift symmetry restoration in the limit $m_a\to 0$ . 

Here, the effects of scattering with an ALP background are examined. It will be seen that the insertion of an axion-like particle scattering into the one-loop QED diagram will generate an apparent electric dipole moment for the electron and proton. This scattering injects parity violation into the diagram and converts the would-be anomalous magnetic moment contribution into an EDM contribution. Although the resulting contribution has some similarities to that found in \cite{Alexander:2017zoh}, it will not contain any of the signs of a fictitious contribution \cite{Berlin:2023ubt}.  While this scattering induced contribution can be modeled by an interaction with a classical background field, it will be seen that the scattering picture will be crucial for validating this contribution\footnote{The origins of the EDM in \cite{Alexander:2017zoh} cannot be similarly interpreted as a scattering, since this scattering would violate conservation of momentum and energy.}.

\section{Scattering of the Background}
\begin{figure}[!t]
\centering
\includegraphics[width=8cm,trim={.7in .55in .7in .7in},clip]{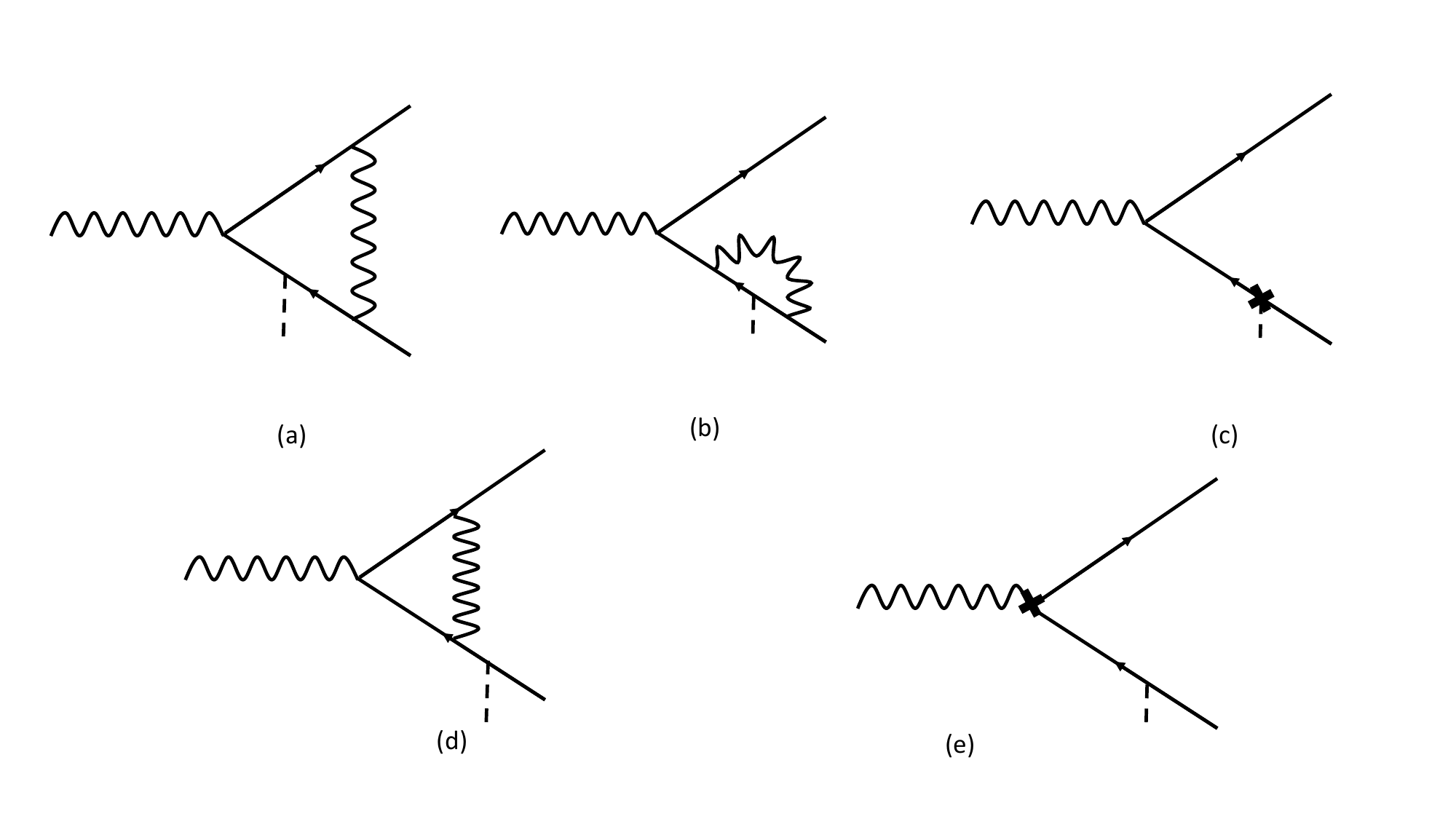}
\caption{\it The Feynman diagrams of the apparent EDM.}
\label{fig:FeynDiag}
\end{figure}

In this section, a missed scattering off an ALP dark matter background is examined. The scattering diagrams in Fig. (\ref{fig:FeynDiag}) will be used to explore the connection between scattering with a collection of particles and a classical field background. This then makes it possible to interpret scattering as an interaction with a classical background field. With this understanding, the diagrams in Fig. (\ref{fig:FeynDiag}) can be interpreted as a time-dependent electric dipole moment.   

Since the diagrams in Fig. (\ref{fig:FeynDiag}) involve a scattering with an ultralight dark matter background, they can only be interpreted as an EDM as long as the experiment is unable to resolve the momentum transfer. Since it will be found that this search method is only competitive for ALP dark matter masses of order $10^{-10}~eV$ or smaller, and the electron and ALP are non-relativistic, the largest possible energy transfer will be 
\begin{eqnarray}
    \frac{\Delta E_e}{E_e}\simeq \frac{m_a}{m_e} \simeq 2.0\times 10^{-16}~.
\end{eqnarray}
%
Thus, to invalidate this search strategy, the ACME II experiment would need to resolve the electron energy to better than about two parts in $10^{16}$, and even more precisely for lighter masses. Clearly, these scatterings will be missed and the process in the diagrams found in Fig. (\ref{fig:FeynDiag}) has the potential to be interpreted as an EDM.  


Scattering matrix elements are generally written in terms of plane wave ingoing and outgoing states. For the processes depicted in Fig.(\ref{fig:FeynDiag}), this would take the form
\begin{eqnarray}
    \langle \bar k_{\bar \psi}|S| k_\psi ,\Delta k_{A^\mu}, p_a\rangle\label{eq:ScatteringSingleParticle}~,
\end{eqnarray}
where $S$ is the scattering matrix and the $k,p$ variables, with the indicative subscripts, specify which particle momenta they represent. The form of the $S$ matrix for the particular processes in Fig. (\ref{fig:FeynDiag}) is
\begin{eqnarray}
   S_{EDM}\supset T \frac{4(-i)^4}{4!}\left(e\int d^4x \bar \psi_i \gamma_\mu \psi_i A^\mu\right)^3\quad\quad\quad \quad \\ \nonumber \times \left(\int d^4x' \bar g_a^i a\bar \psi_i i\gamma_5 \psi_i\right) ~,
\end{eqnarray}
where repeated indices are not summed and $T$ is for time ordering. Within the matrix element in Eq. (\ref{eq:ScatteringSingleParticle}) is contained the following matrix element for the ALP
\begin{eqnarray}
    \langle 0|a(x)|p\rangle = e^{ipx}~.
\end{eqnarray}
Since this matrix element assumes no background and the particle is in a momentum eigenstate, it gives a plane wave. For experiments that are unable to resolve the position of the ALP at the level of a de Broglie wavelength, this plane wave approximation is justified, but is merely an approximation to simplify calculations.  

When considering matrix elements between $N$ and $N-1$ particle states, it is easier to use states composed of wave packets\footnote{Furthermore, since the process considered here is the scattering of a fermion off a physical background particle, it must be a wave packet and not the projection of a wave packet onto a plane wave.}. The wave packet nature of the particle can be incorporated into the scattering matrix element by modifying the in-going state (or likewise an outgoing state) as follows 
\begin{eqnarray}
    | p_i\rangle  \to |a\rangle = a^\dagger_{a}|0\rangle = \int \frac{d^3p_i}{(2\pi)^3} \frac{\tilde  a(p_i)}{\sqrt{2E_{p_i}}}|p_i\rangle ~,
\end{eqnarray}
where 
\begin{eqnarray}
    a^\dagger_a=\int \frac{d^3p}{(2\pi)^3} \tilde a(p)a^\dagger(p)~,
\end{eqnarray}
$a^\dagger(p)$ is the creation operator for the ALP, and $\tilde  a(p)$ contains the information about the momentum spread of the particle. The function $\tilde a(p)$ must satisfy
\begin{eqnarray}
    \int \frac{d^3p}{(2\pi)^3} |\tilde  a(p)|^2 =1 ~,
\end{eqnarray}
and, if it describes dark matter, is only non-trivial for momentum of order $p_0=m_a v_a$, with $v_a\simeq 10^{-3}$. The normalization condition on $\tilde a(p)$ then imposes a scaling behavior of 
\begin{eqnarray}
    \tilde  a(p)\sim \frac{1}{p_0^{3/2}}~.\label{eq:BarAScal}
\end{eqnarray}
With this generalization of the in-going ALP state, the relevant matrix element for scattering off a single particle becomes 
\begin{eqnarray}
 \langle \bar k_{\bar \psi}|S_{EDM}| k_\psi ,\Delta k, a\rangle ~,
\end{eqnarray}
and the ALP portion of this matrix element is now
\begin{eqnarray}
 \langle 0|a(x)|a\rangle = \int \frac{d^3p_i}{(2\pi)^3} \frac{\tilde a (p_i)}{\sqrt{2E_{p_i}}} e^{ipx}=\bar a(x)~.\label{eq:baraofx}
\end{eqnarray}
As is clear from this equation, the state has been modify to the physical wave function of the particle. By examining the scaling behavior of $\tilde a(p)$ in Eq. (\ref{eq:BarAScal}), it becomes clear that  
\begin{eqnarray}
    \bar a(x) \sim \frac{p_0^{3/2}}{m_a^{1/2}} ~, \label{eq:abarScale}
\end{eqnarray}
if the particle is assumed to be non-relativistic. Since $p_0\sim m_a v_a$, $\bar a(x)$ is proportional to $m_a$ and the scattering amplitude vanishes in the limit $m_a\to 0$.

To discuss scattering off a background, the matrix element needs to be modified to a transition between $N$ and $N-1$ particle states. This gives the modification 
\begin{eqnarray}
    \langle \bar k_{\bar \psi}|S_{EDM}| k_\psi ,\Delta k, a\rangle\quad\quad\quad\quad\quad\quad \quad\quad\quad\quad \\ \nonumber \to \langle \bar k_{\bar \psi},a(N)|S_{EDM}| k_\psi ,\Delta k, a(N-1)\rangle
\end{eqnarray}
where 
\begin{eqnarray}
    | a(N)\rangle  = \frac{\left[a_a^\dagger\right]^N}{\sqrt{N!}}|0\rangle  ~.
\end{eqnarray}
With this further generalization, the ALP matrix element now takes the form,
\begin{eqnarray}
\langle a_i(N_i-1)|a(x') |a_i(N_i)\rangle =\sqrt{N_i}\bar a_i(x')\label{eq:sqrtNiax}~,
\end{eqnarray}
where $N_i$ is the number of identical particles in the background. 

Before addressing how to incorporate the quasi-coherent nature of the ALP background, the scaling of this scattering amplitude for $m_a\to 0$ is examined. If $N_i$ is fixed, this matrix element scales exactly the same as the single particle scattering, Eq. (\ref{eq:abarScale}), and decouples in the $m_a\to 0 $ limit. However, the scaling is modified for a background with energy density fixed to the dark matter measured value. For fixed energy density, the number of particles, $N_i$, is roughly
\begin{eqnarray}
    N \simeq  \frac{\rho_{DM}}{m_a \frac{\frac{4\pi}{3} p_0^3}{(2\pi)^3}}.
\end{eqnarray}
The important thing to note in this expression is that the number of particles scales with $m_a^{-4}$, since $p_0=m_a v_a$. Combining this expression with that in Eq. (\ref{eq:abarScale}), it is seen that the total scaling of the ALP matrix element with $m_a$ will be 
\begin{eqnarray}
    \bar a(x) \sqrt{N_i} \sim \frac{\sqrt{\rho_{DM}}}{m_a}~.
\end{eqnarray}
This is consistent with the typical scaling found for the classical field approximation of an ALP background, as discussed below. Thus, although the decoupling of this matrix element in the $m_a\to 0$ limit is obscured by the occupation number scaling, it does exhibit decoupling.

The quasi-coherent nature of the background can now be incorporated into this scattering calculation by parsing the background into coherent states with $N_i$ particles and then summing over the scattering rate to each state. This then gives a total ALP matrix element of 
\begin{eqnarray}
 && \sum\limits_i \langle a_i(N_i-1)|a(x') |a_i(N_i)\rangle \\ \nonumber &&\quad\quad\quad\quad  =\sum\limits_i \sqrt{N_i}\bar a_i(x')=\bar a_{T}(x') \label{eq:aTOT}~.
\end{eqnarray}
Using $\rho_{DM}=\frac{1}{2} m_a^2 a^0$, the amplitude of the individual oscillations of the axion fields can be approximated as\cite{Foster:2017hbq}
\begin{eqnarray}
    \bar a_i (x)\simeq \frac{\sqrt{2\rho_{DM}/N_a}}{m_a} \cos\left[m_a\left(1+\frac{v_i^2}{2}\right)t +\phi_i\right]~,
\end{eqnarray}
If the different coherent states with $N_i$ particles are properly summed over, it is found that \cite{Foster:2017hbq}  
\begin{eqnarray}
    \bar a_{T} (t)=\sum\limits_i\alpha_j\frac{\sqrt{\rho_{DM}}}{m_a} \sqrt{\frac{N^j_a}{N_a}} \cos\left[m_a\left(1+\frac{v_j^2}{2}\right)+\phi_j\right]~,~~\label{eq:aTOTSUM}
\end{eqnarray}
where $\alpha_j$ is a random variable sampled from the Rayleigh probability distribution \footnote{See \cite{Foster:2017hbq} for more details.}. 

Since the $S$ matrix considered here only has a single ALP in it, the operation of the $S$ matrix on the $N$ and $N-1$ states leads to an effective scattering matrix of 
\begin{eqnarray}
   S_{EDM}\supset T \frac{1}{3!}\left(e\int d^4x \bar \psi_i \gamma_\mu \psi_i A^\mu\right)^3\quad\quad\quad \quad \\ \nonumber \times \left(\int d^4x' \bar g_a^i \bar a_{T}(x)\bar \psi_i i\gamma_5 \psi_i\right) ~.
\end{eqnarray}
Thus, the calculation of scattering is equivalent to scattering of the quasi-coherent classical background field found in Eq. (\ref{eq:aTOTSUM}).

The quasi-coherent nature of the ultralight dark matter background leads to fluctuations in the amplitude of the ALP fields oscillation amplitude. This fluctuation weakens ALP dark matter constraints as shown in \cite{Centers:2019dyn}. Instead of performing a detailed analysis of this effect here, the constraints are just divide by $2.7$ for all bounds as suggested in \cite{Centers:2019dyn}.

\section{Electric Dipole Moment}
Here, the diagrams in Fig. (\ref{fig:FeynDiag}) are calculated using on-shell renormalization scheme for the following set of interactions,
\begin{eqnarray}
  {\cal L}_I=\frac{g_a^i}{2m_i}\left(\partial_\mu a\right)\bar\psi_i\gamma^\mu\gamma_5\psi_i+\bar g_a^i a\bar\psi_i i\gamma_5\psi_i~, \label{eq:aInt}
\end{eqnarray}
where $\psi_i$ can be either the electron or the proton. 

Initially, $\bar g_a^i\ne 0$ and $g_a^i=0$ will be discussed. For this choice of couplings, Fig (\ref{fig:FeynDiag}b) and (\ref{fig:FeynDiag}c) give a contribution which is infrared divergent and of the form
\begin{eqnarray}
    iM_{(b+c)_1}=F(m_i,m_\gamma)\bar u(\bar k)\gamma_\mu\gamma_5 u(k)~,
\end{eqnarray}
where $u(k)$ is the fermion spinor and $F(m_i,m_\gamma)$ is a function of the fermion mass, $m_i$, and photon mass, $m_\gamma$, which is used to regulate the diagram. There is a nearly identical contribution from diagrams with an ALP scattering off the other fermion leg. These contributions differ only in the order of $\gamma_5$ and $\gamma_\mu$. The sum of these contributions then vanish,
\begin{eqnarray}
    iM_{(b+c)_{tot}}=F(m_i,m_\gamma)\bar u(\bar k)\left[\gamma_\mu\gamma_5+\gamma_5\gamma_\mu\right] u(k)=0~.~
\end{eqnarray}
Importantly, this cancels the associated IR divergences.  

Next, the diagrams in Fig. (\ref{fig:FeynDiag}a), (\ref{fig:FeynDiag}d), and (\ref{fig:FeynDiag}e) are calculated. These diagrams must be calculated together, since separately they all diverge. To simplify the calculation, the diagram in Fig. (\ref{fig:FeynDiag}d) is expanded in terms of  Fig. (\ref{fig:FeynDiag}a) as follows,
\begin{eqnarray}
    iM_d=iM'_d -iM_a +{\mathcal O}(p)~.
\end{eqnarray}
where $iM_d$ is the full contribution from the diagram in Fig. (\ref{fig:FeynDiag}d) and $iM_d'$ is the remaining part after the contribution in Fig. (\ref{fig:FeynDiag}a) is removed.  This means the diagram in Fig. (\ref{fig:FeynDiag}a) need not be calculated. The diagram in Fig. (\ref{fig:FeynDiag}e) contains the QED counter term. This QED counter term can be found in many quantum field theory text books and inserted into this diagram. The resulting contribution of these three diagrams, to leading order in $p$, is then
\begin{eqnarray}
iM_{tot}&=&  ie \frac{\alpha}{2\pi} \bar u_i(\bar k) \left[ \frac{ \left[\Delta \slashed k,\gamma_\mu\right]\gamma_5}{4m_i^2}\right.  \label{eq:TotEDM}    \\  &+& \left.\frac{\gamma_5\slashed p \left[\Delta \slashed k,\gamma_\mu\right]  }{8\bar k\cdot p m_i}+\frac{ \left[\Delta \slashed k,\gamma_\mu\right]\slashed p \gamma_5}{8k\cdot p m_i }\right] u_i(k)\bar a_{T}(p) ~, \nonumber 
\end{eqnarray}
and $\bar a_{T}(p)$ is the Fourier transform of Eq. (\ref{eq:aTOTSUM}) and $\bar u(\bar k),u(k)$ are the fermion spinors of the in-going and outgoing states respectively. The terms in the second line are ignored, since they only contribute to the off-diagonal part of the Dirac equation and are highly suppressed in the non-relativistic limit. The contribution from the first line of Eq. (\ref{eq:TotEDM}) is then mapped back to a Lagrangian of the form
\begin{eqnarray}
    {\mathcal L} \supset e\frac{\alpha}{2\pi} \frac{\bar g_a^i\bar a_{T}(x)}{4m_i^2}  \bar \psi_i\sigma_{\mu\nu}i\gamma_5\psi_i F^{\mu\nu}~. \label{eq:EDMbarg}
\end{eqnarray}
where $\alpha$ is the fine structure constant. Importantly, this contribution to the EDM is relativistic, avoiding another one of the indicators of a fictitious contribution to an EDM.

Since the coupling $\bar g_a^i$ can be rotated away (to leading order) by taking the field redefinition $\psi \to e^{i\gamma_5\bar g_a^i \frac{ a(x)}{2m_i}}\psi $, the contribution in Eq. (\ref{eq:EDMbarg}) can also be rotated away. However, this rotation generates the coupling $g_a^i$.

To be assured that the contribution in Eq.(\ref{eq:TotEDM}) is indeed physical, the diagrams in Fig. (\ref{fig:FeynDiag}) are recalculated with $g^i_a\ne 0$ and $\bar g_a^i= 0$. In this form of the theory, the axion-like particle's coupling is proportional to its momentum $p$. Naively, this makes it appear the EDM contribution is suppressed by the axion-like particle's momentum.  However, this is only true for the diagram in Fig. (\ref{fig:FeynDiag}a). 

The remaining diagrams are not suppressed due to the fermion propagator, which takes the following form,
\begin{eqnarray}
    \frac{\slashed k \pm \slashed p+m_i}{(k\pm p)^2-m_i^2}= \frac{\slashed k+m_i}{\pm 2k\cdot p}+{\mathcal O}(p^0)\label{eq:PropOnShell}~,
\end{eqnarray}
where $k$ is the on-shell momentum of the fermion and satisfies $k^2-m_i^2=0$. The ${\mathcal O}(p^0)$ term to the right of the equality can be neglected, since the momentum-dependent coupling leads to additional $p$ suppression, making this contribution negligible. This greatly simplifies the calculation, since the numerator of the propagator can be replaced by on-shell Dirac spinors that satisfy the following relationship 
\begin{eqnarray}
    \sum u^s (k)\bar u^s (\bar k) = \slashed k+m_i~. 
\end{eqnarray}
This allows the propagator in Eq. (\ref{eq:PropOnShell}) to be written as 
\begin{eqnarray}
    \frac{\sum u^s(k)\bar u^s(k)}{\pm 2p\cdot k} ~.
\end{eqnarray}

With this simplification, the loops of the remaining diagrams in Fig(\ref{fig:FeynDiag}) have on-shell spinors for the in-going and outgoing fermions. This means the quantum electrodynamics (QED) loop-corrected vertex in Fig. (\ref{fig:FeynDiag}d) can be calculated independently of the rest of the diagram and then inserted back into the diagram. This is equivalent to replacing the QED corrected vertex with the leading order contribution to the anomalous magnetic moment. The sum of the remaining diagrams, Fig. (\ref{fig:FeynDiag}b) and Fig. (\ref{fig:FeynDiag}c), is found to vanish to leading order in $p$. When summed up, this gives exactly the same contribution as that found in Eq. (\ref{eq:TotEDM}). Thus, the contributions to the electron and proton EDM from the couplings $g_a^i$ and $\bar g_a^i$ are identical.


If the derivative interaction, $g_a^i$, had led to a suppressed contribution, this would have cast doubt on whether the calculated contribution involving the coupling $\bar g_a^i$ was indeed physical. However, since they are identical, it strongly supports the conclusion that a missed fermion scattering with a background axion-like particle results in an apparent EDM. It has now been shown that this contribution avoids all the telltale signs of a fictitious contribution to an EDM. 

\section{Resulting Constraints}
\begin{figure}[tp!]
    \centering
    \includegraphics[width=.9\linewidth,trim=1.25cm 0 1.3cm 0, clip]{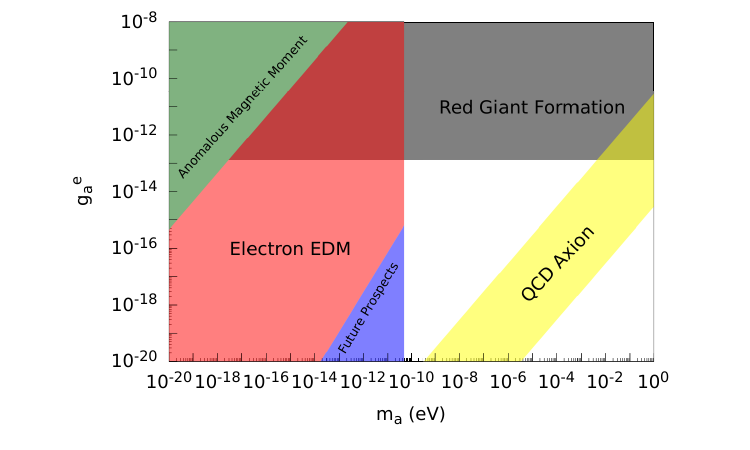}
    \caption{The gray shaded region shows previous constraints from red giant observations\cite{Capozzi:2020cbu}. The green shaded region is excluded by measurements of the anomalous magnetic moment \cite{Arza:2023wou,Evans:2024dty,Fan:2022eto}. The yellow shaded area is a very liberal band for the QCD axion. The red shaded region is the new constraints coming from ACME II's measurement of the electron EDM, with $d_e< 1.1\times 10^{-29}~e~cm$ \cite{ACME:2018yjb} applied below 100 kHz. The blue shaded region is future prospects for detection of ultralight dark matter by electron EDM experiments.}
    \label{fig:NewConstraints}
\end{figure}

The correction to the Lagrangian found in Eq. (\ref{eq:EDMbarg}) gives an EDM of 
\begin{eqnarray}
    d_i=e\frac{\alpha}{2\pi} \frac{g_a^i\bar a_{T}(t)}{m_i^2} \label{eq:dipredicted}~.
\end{eqnarray}
Due to the homogeneity of the dark matter background, the space dependence of $a_{T}(x)$ has been ignored in the above expression. As mentioned earlier, the stochastic nature of the background field weakens the constraint by about a factor of 2.7 compared to a completely coherent background \cite{Centers:2019dyn}. I will apply this factor to the predicted value and consider the ALP field here to be effectively coherent. In this case, the EDMs for the electron and proton can then be approximated as 
\begin{eqnarray}
    d_e= 7.03\times 10^{-30} (e~cm)\left(\frac{g^e_a}{10^{-17}}\right)\quad \quad\quad\quad\\
    \left(\frac{10^{-10}~{\rm eV}}{m_a}\right)\cos(m_a t) \nonumber ~,\\
   d_P= 2.09\times 10^{-25} (e~cm)\left(\frac{g^P_a}{10^{-11}}\right)\quad \quad\quad\quad\\
    \left(\frac{10^{-15}~{\rm eV}}{m_a}\right)\cos(m_a t) ~,
    \end{eqnarray}
where the current upper bound is 
\begin{eqnarray}
&&d_e<1.1 \times 10^{-29} e~cm~,\label{eq:deConstraint} \text{\cite{ACME:2018yjb}},\\
&&d_P< 2.1\times 10^{-25} e~cm~, ~\text{\cite{Sahoo:2016zvr}}~. \label{eq:dpConstraint}     
\end{eqnarray}

For the electron EDM, which is measured very precisely in ACME II, a time-dependent EDM with a period of order one second or longer can be measured with roughly the same accuracy as a constant EDM. This means that for $m_a< 6.58\times 10^{-16}~{\rm eV}$ Eq. (\ref{eq:deConstraint}) can be applied. This allows us to take the coefficient of the time-dependent electron EDM and compare it to Eq. (\ref{eq:deConstraint}) to get
\begin{eqnarray}
    g_a^e< 1.56\times 10^{-27} \frac{m_a}{10^{-20}~{e~cm}}~.
\end{eqnarray}
This gives a constraint that is over eleven orders of magnitude stronger than the anomalous magnetic moment constraints\cite{Arza:2023wou,Evans:2024dty} for $m_a=10^{-20}$ eV. For $m_a> 6.58\times 10^{-16}~{\rm eV}$, the constraint begins to weaken. Furthermore, it is unlikely that the ACME II experiment can constrain time-dependent EDMs with variations that occur faster than $100~kHz$\footnote{From private discussion with ACME II experimentalist.}. Thus, the constraint completely disappears if $m_a\gtrsim 6.58\times 10^{-11}~{eV}$. At this edge where sensitivity disappears, the constraint on the electron EDM is taken to be two orders of magnitude weaker than that shown in Eq. (\ref{eq:deConstraint})\footnote{Where and how much the constraint weakens due to the time-dependence of the EDM is currently under study by the ACME II collaboration. Based on discussion with a member of the group, the constraints placed in Fig. (\ref{fig:NewConstraints}) are fairly conservative.}. To account for this weakening, the constraint is given some mass dependence to extrapolate from $m_a=6.58\times 10^{-16}~{\rm eV}$ to $m_a\sim 6.58\times  10^{-11}~{\rm eV}$ as follows
\begin{eqnarray}
    d_e=1.1\times 10^{-29}~e~cm \left(\frac{m_a}{6.58\times 10^{-16}~{\rm eV}}\right)^{2/5} ~.
\end{eqnarray}
This then leads to a constraint on the ALP coupling to an electron of 
\begin{eqnarray}
    g_a^e<1.85\times 10^{-22} \left(\frac{m_a}{10^{-15}~{\rm eV}}\right)^{\frac{6}{5}}~,
\end{eqnarray}
for $m_a\gtrsim 6.58\times 10^{-16}$. 

A figure with the new constraints on axion-like particles dark matter coupling to the electron and a comparison to previous constraints can be found in Fig. (\ref{fig:NewConstraints}).  As is clearly seen from the red shaded region, this method of searching for ALP dark matter vastly outperforms any previous methods.  The constraint is so strong that only masses above about $m_a=10^{-15}~{\rm eV}$ are seen for the range of $g_{ae}$ presented in the figure. Unless the electron ALP coupling is significantly smaller than $(4\pi)^2m_eg_{a\gamma\gamma}$, where $g_{a\gamma\gamma}$ is the ALP photon coupling, this red-shaded region will exclude some of the preferred dark matter range for axion-like particles with different early universe evolution \cite{Blinov:2019rhb,Liang:2024vnd}. Furthermore, the sensitivity of these types of experiments has the potential to be improved up to six orders of magnitude \cite{Vutha:2017pej,Kozyryev:2017cwq,Fitch:2020jil,Chupp:2017rkp}  for a static EDM.  In Fig. (\ref{fig:NewConstraints}), I have put the future prospects in blue where I have assumed that the sensitivity of these experiments to time-dependent EDMs is similar to the conservative estimates for ACME II above. If the time dependence could be probed beyond $10^{-10}$ eV, this method could potentially probe the QCD axion. 

Now, the proton EDM and its constraints on the proton coupling to the axion are considered in the light mass range, where the calculated contribution can be considered static. The relevant limit comes from applying the bound in Eq. (\ref{eq:dpConstraint}) to the contribution in Eq. (\ref{eq:dipredicted}). This leads to the following constraint 
\begin{eqnarray}
    g_a^P< 1.01\times 10^{-16} \left(\frac{m_a}{10^{-20}~{\rm eV}}\right)~.
\end{eqnarray}
Since, the period of oscillation at the mass scale $10^{-20}~{\rm eV}$ is of order a day, this should be a viable measurement at this mass scale.  The experiment takes measurements of the Hg EDM over a period of one second. Thus, an oscillation with a period of one day should be easily resolved\footnote{At what scale this bound becomes invalid is unclear and is left for the experimentalist to answer later.}. Since the current bound is $g_a^P\lesssim 6\times 10^{-10}$ \cite{Lella:2023bfb}, this is an improvement of nearly seven orders of magnitude.

\section{Conclusions}
If dark matter is ultralight, its background occupation number is enormous. With so many particles in the background, it is nearly impossible to avoid scattering with dark matter. However, because the momentum transfer of these scatterings is so small, the energy transfer of a single scattering with the background will undoubtedly be missed in any experimental setup. If the occupation number of the background is large enough, like it is for ultralight dark matter, the collective effect of scattering off the background can be non-trivial. Here, it was shown that one such effect is to inject parity violation creating a time-dependent relativistic EDM for the electron and proton. The total effect of this scattering process with the background dark matter can be modeled by a classical background field. The resulting EDM is proportional to this classical background field and is the same for derivative and non-derivative couplings. The basis independence of the result and the decoupling in the $m_a\to0$ limit, for a fixed occupation number, strongly support the physicality of this contribution.

The resulting EDM of the electron from scattering with the background is so significant that it leads to new constraints in the mass range $m_a=10^{-11}~{\rm eV}$ \textemdash $10^{-20}~{\rm eV}$, which are as much as eleven orders of magnitude stronger than previous constraints. The resulting EDM of the proton is also quite large, with the new constraints almost seven orders of magnitude stronger than previous constraints for $m_a=10^{-20}~{\rm eV}$.  Thus, the influence of scattering with this ultralight particle background can have a significant effect on precision measurements and leads to new discover paths for ultralight dark matter.




\section*{Acknowledgement}
I would like to thank Satoshi Shirai, Tsutomu T. Yanagida, Xing Fan and the ACME II Collaboaration for useful discussions regarding to this work. 

\bibliography{MissedScatterEDM}

\end{document}